# *From Values to Frameworks: A Qualitative Study of Ethical Reasoning in Agentic AI Practitioners*

Theodore Roberts[1], Bahram Zarrin[2]

[1] Dartmouth College, Ethics Institute  [2] Microsoft Research Hub

# Abstract

Agentic artificial intelligence systems are autonomous technologies capable of pursuing complex goals with minimal human oversight and are rapidly emerging as the next frontier in AI. While these systems promise major gains in productivity, they also raise new ethical challenges. Prior research has examined how different populations prioritize Responsible AI values, yet little is known about how practitioners actually reason through the trade-offs inherent in designing these autonomous systems. This paper investigates the ethical reasoning of AI practitioners through qualitative interviews centered on structured dilemmas in agentic AI deployment. We find that the responses of practitioners do not merely reflect value preferences but rather align with three distinct reasoning frameworks. First is a Customer-Centric framework where choices are justified by business interests, legality, and user autonomy. Second is a Design-Centric framework emphasizing technical safeguards and system constraints. Third is an Ethics-Centric framework prioritizing social good and moral responsibility beyond compliance. We argue that these frameworks offer distinct and necessary insights for navigating ethical trade-offs. Consequently, providers of agentic AI must look beyond general principles and actively manage how these diverse reasoning frameworks are represented in their decision-making processes to ensure robust ethical outcomes.

• Social and professional topics → Codes of ethics

# Introduction

Agentic AI is poised to fundamentally reshape the global economy and the nature of work. Unlike traditional AI assistants that merely respond to user commands, agentic AI is designed to autonomously pursue complex goals—coordinating across multiple tools, adapting over time, and engaging users through multimodal interaction strategies.[1] Along with its transformative effects on work come many challenging dilemmas, meaning that RAI is becoming a growing focus area for applied ethics.[2]

The pace of change and the complexity of the ethical problems that occur in designing and implementing agentic AI means that AI practitioners will often face tradeoffs between RAI values where regulatory systems, legal requirements, company policies, and moral intuitions do not provide clear answers. By AI practitioners, I mean professionals involved in the design, development, marketing, and/or sales of agentic AI. Even as AI ethics guidelines converge around a finite set of values, practitioners are in many ways left to decide for themselves the "right" thing to do.[3]

---

[1] Lee, J. (2025, March 3). *The evolution of AI: From AlphaGo to AI agents, physical AI, and beyond*. MIT Technology Review. https://www.technologyreview.com/2025/02/28/1112530/the-evolution-of-ai-from-alphago-to-ai-agents-physical-ai-and-beyond/

[2] See, e.g., Gabriel, I., Manzini, A., Keeling, G., Hendricks, L. A., Rieser, V., Iqbal, H., Tomašev, N., Ktena, I., Kenton, Z., Rodriguez, M., El-Sayed, S., Brown, S., Akbulut, C., Trask, A., Hughes, E., Bergman, A. S., Shelby, R., Marchal, N., Griffin, C., . . . Manyika, J. (2024, April 24). *The ethics of advanced AI assistants*. arXiv.org. https://arxiv.org/abs/2404.16244

Prior work has shown that value preferences can vary based on the identity of the decision maker and between AI practitioners and the general population, highlighting the importance of paying attention to who gets to define "Responsible AI." Building on those findings, this work seeks to understand variations in approaches used by AI practitioners.

In this work, we present the results of qualitative research into the ways AI practitioners apply ethical principles. This work asks participants to reason through three fact scenarios relating to potential ethical trade-offs associated with business applications of agentic AI. Using thematic analysis to extract results from this qualitative data, this paper's results provide evidence that participants apply three distinct frameworks of thinking about these problems in consistent patterns. By framework I mean the way in which the participants interpret the situations, reason through decisions, and justify their priorities. Based on those patterns, on average, respondents will approach ethical trade-offs using a framework that places a larger emphasis on customer-centric, design-centric, or ethics-centric considerations. We observed that each of these approaches provided distinct and valuable insights that can usefully inform reasoned ethical decision-making.

If valid, these findings suggest that providers of agentic AI should consider not just the identity of the people making decisions, but also their approach to applying RAI principles. The qualitative approach in this study compliments prior empirical approaches[4]. However, as we elaborate in the discussion, the limited sample size and study design should be expanded and replicated to develop more robust support for the findings and recommendations in this paper.

# Related Work

Previous research by Microsoft Research Hub[4] compared how different groups prioritize ethical values for RAI. Jakesch et al. (2022)[4] surveyed three distinct groups—AI practitioners who professionally develop AI, crowdworkers who previously contributed to the training of AI models through things like data labeling, and a representative sample of the U.S. public—on how they prioritize 12 RAI values. Their findings revealed clear differences: While the general public and crowdworkers emphasized values like safety, privacy, and performance, AI practitioners were more likely to prioritize fairness, inclusiveness, and dignity. Interestingly, practitioners also tended to rate RAI values as less important overall. As such, they concluded that AI practitioners are acting with a different set of values in mind than the general population, which poses a risk to the future responsible development of AI.

| Fairness | Privacy | Sustainability |
|---|---|---|
| Inclusiveness | Safety | Social good |
| Dignity | Performance | Accountability |
| Transparency | Human autonomy | Solidarity |

---

[3] Some may argue that people are always left to decide ethical questions without the benefit of any moral framework that will give them the "right" answer. See, e.g., Juan Paul Sartre's parable of the soldier who must decide whether to stay home with his mother or join the French Free Forces fighting in World War II. Sartre, Jean-Paul. 2003. *Being and Nothingness*. Translated by Hazel Estella Barnes. 2nd ed. London, England: Routledge.

[4] Jakesch, M., Buçinca, Z., Amershi, S., & Olteanu, A. (2022). How different groups prioritize ethical values for responsible AI. 2022 ACM Conference on Fairness, Accountability, and Transparency, 310–323. https://doi.org/10.1145/3531146.3533097

**Table 1: Values identified by Jakesch et al. (2022) for prioritization as the most important for AI systems by survey participants. See Appendix 1 for provided definitions.**

While their survey-based approach offered larger-scale empirical insights, its structure provided limited information to understand how participants' reason through ethical dilemmas and why they prioritize certain values over others. For example, participants were given an example question like:

> *"A movie streaming company uses an AI system that scans users' data to predict which other movies they would enjoy seeing. A list of recommended movies is automatically shown to thousands of users based on the output of this AI system.*
>
> *The developers realize that making the system's predictions possibly accurate (ensuring performance) may require the collection of additional sensitive data (reducing privacy). Should they prioritize performance or privacy?*
>
> *Definitely performance | Probably performance | Undecided | Probably privacy | Definitely privacy."* [4]

While the study captured participants' responses, it did not reveal their reasoning—such as the values, stakeholders, or ethical frameworks they considered. This research aims to fill that gap by exploring the rationale behind participants' choices and categorizing their reasoning.

To build on and complement their work, this study used interviews to explore the reasoning behind AI practitioners' decisions in the context of agentic AI systems. Using this approach, the goal was to replicate the previous study's results and explore why people respond the ways they do.

This process revealed a more complex understanding of practitioners' responses: It is about more than just the values they have, it is about their framework of thinking. By shifting the focus from value preferences to reasoning processes, this work seeks to build on the findings of Jakesch et al[4] and extend into new territory—probing not just what values people claim to prioritize, but how they justify those priorities in the complex and novel context of agentic AI.

# Methodology

This paper uses qualitative data from 30-minute interviews with AI practitioners (n=11) involved in developing, marketing, or selling agentic AI to determine the ethical values they used to think about specific agentic AI ethical scenarios. Each interview consisted of asking the participants to reason through three agentic AI related scenarios, ensuring that they both gave their opinion on what should be done as well as their reasoning behind it.

Each scenario was designed to put several of the 12 RAI values established in the previous Research Hub paper in conflict, requiring participants to indicate which values they found the most salient (as indicated in Figure 2, below). These scenarios are based on agentic AI systems operating within manufacturing to keep these factors constant. For each scenario, the study presented participants with an ethical dilemma that consisted of multiple parts that at first were designed to pull intuitions in one direction, then advance the severity of the situation to pull them in a different direction. The scenarios were based on a combination of real-world situations and theoretical examples.[5]

---

[5] Note, the scenarios are (a) hypotheticals, not based on any actual products or customers; and (b) designed for AI professionals providing business applications to commercial customers. The scenarios were designed so as not to require the understanding or application of any legal or regulatory requirements.

To identify the frameworks participants were employing, after they responded, each was prompted with follow up questions designed to surface their reasoning behind each decision. This process encouraged them to articulate the rationale and underlying assumptions guiding their judgments. Through this process, the conceptual framework that the interviewees were using became clearer. All responses were recorded and noted using a standardized template that captured not only their choices but also their justifications (shown in Appendix 3), allowing for consistent comparison and thematic coding across participants. Below is a table summarizing the basic dilemma of each scenario as well as the primary RAI values that they were designed to place at odds:

| Scenario Summary | Primary RAI Values |
|---|---|
| Testing the extent AI agents should push the limits<br>1.1 Beating the system<br>1.2 Using aggressive (but legal) tactics<br>1.3 Holding AI to a higher standard than humans | Accountability, Dignity, Fairness, Performance, Privacy, Safety, Social Good |
| Scheduling optimization (testing workers' rights and the extent to which AI can be used to improve productivity)<br>2.1 Possible self-reinforcing preferences<br>2.2 Use for impactful decisions<br>2.3 Disparate impact on groups<br>2.4 Allegations of bias and discrimination | Autonomy, Fairness, Performance, Privacy, Social Good, Solidarity |
| Sustainable procurement (testing the extent to which AI decisions should adjust for externalities and how shifting the participants' role may impact decision-making)<br>3.1 Customer complaint (in your role)<br>3.2 Default setting and configurability (as decision maker)<br>3.3 Integrating social goods into AI (as a citizen) | Performance, Social Good, Sustainability, Transparency |

**Table 2: Scenarios used with AI practitioners and the primary RAI values expected to be part of the reasoning and trade-offs. See Appendix 2 for the full scenarios.**

After the interviews, the study used thematic analysis to determine which of the 12 RAI values the participants prioritized, and the reasons given to justify their decisions. Thematic analysis is a way to analyze qualitative data, such as from interview feedback from users.[6] Applied to this work, the thematic analysis used the following steps: (1) transcribe the recorded interviews; (2) identify meaningful ethical reasoning or explicit value references in the transcript and assign initial codes based on salient features; (3) group similar codes and identify patterns of meaning across participants to organize these codes into candidate themes; (4) review the initially assigned codes with special attention to ensure that they accurately reflect the nuances of participants' reasoning, not just frequency of occurrence, and refine when needed; (5) review the candidate themes and refine iteratively, comparing them back to the full dataset to ensure internal coherence and external distinction; and (6) assign names to the identified themes to describe the primary concepts coded within each theme (as detailed in Fig. 3, below).

---

[6]Jason, L., & Glenwick, D. (2016). Handbook of Methodological Approaches to Community-based Research: Qualitative, quantitative, and mixed methods. Oxford University Press, 33-41. https://repository.unar.ac.id/jspui/bitstream/123456789/5294/1/Handbook%20of%20methodological%20approaches%20to%20community-based%20research%20%20qualitative%2C%20quantitative%2C%20and%20mixed%20methods%20by%20Leonard%20A.%20Jason%2C%20David%20S.%20Glenwick%20%28z-lib.org%29.pdf#page=54.

# Results

The results from these interviews were surprising at first. While a few of the participants' responses aligned well with the value-based ethics framework that were expected based on prior research, others' answers resulted in very low correlation with any of the 12 RAI values. In fact, most participants very rarely relied on the hypothesized value-based reasoning established in the previous research paper. Many AI practitioners approached the questions not through the lens of ethics or RAI values, but instead through perspectives related to legal standards, company policy, or business practices related to their roles. This divergence highlighted that, while ethics may underpin some decisions, there are several frameworks that the AI practitioners use to approach these problems, which lead them to different kinds of answers.

The thematic analysis of the frameworks behind their reasoning returned clearer patterns. Following the process described above in the methodology section, this study identified three frameworks to describe how participants constructed their arguments about agentic AI: Customer-Centric, Design-Centric, and Ethics-Centric. In Figure 3 below, each framework is defined and reinforced with textual examples from the interviews.

| Framework | Definition and Examples |
|---|---|
| Customer-Centric | Characterized by opinions that emphasize legality, compliance, and customer autonomy. Participants who used this framework often viewed AI tools as neutral instruments, with ethical responsibility resting on the user. For example, Participant 4 (Software Engineer) stated, "You can't put any type of responsibility on the tool… it's the people using the tool that decide how to use it," and compared AI to everyday tools, saying, "If I use OneNote to map a robbery… is it OneNote's fault?" This perspective reflects a belief that an AI provider's role is to provide trustworthy, compliant tools, while customers are responsible for how they are used. Participant 2 (Product Manager) echoed this view, noting, "We should do that in a way that aligns with our values… but in a lot of cases our customers are just closer to the decision-making process." These views also highlight a focus on business outcomes, with Participant 2 (Product Manager) adding, "Revenue is an important factor of running a business, but so is customer satisfaction and customer long term satisfaction." |
| Design-Centric | Characterized by opinions that focus on technical safeguards, transparency, and human oversight, often closely aligning with RAI policy. Participants in this category emphasized the importance of building systems that can adapt, learn from feedback, and avoid harm. Some examples of responses I categorized as Design-Centric include Participant 3 (Engineering Manager) explaining that our "agent should be able to adapt its behavior based on the feedback it receives," and stressed the need for internal checks, saying, "there should definitely be a mechanism in the agent itself to detect whether the strategy it's using is not good." Similarly, Participant 8 (Software Engineer) emphasized the role of human judgment, stating, "A human needs to review this information and make an educated decision based on that," and "The agent can suggest, give supporting data, but it needs to be a human |

| | | |
|---|---|---|
| | being that makes the final decision." These perspectives reflect a commitment to responsible design through modularity, configurability, and shared responsibility between the AI provider and its customers. | |
| Ethics-Centric | Characterized by opinions that prioritize moral responsibility, fairness, and societal well-being and going beyond what is required by either law or company policy. While this category's existence does not mean that there does not exist ethical arguments for the responses of participants using the prior to frameworks, it involves the explicit mention of concepts of morality or the greater good in a way that the other categories do not. Participants using this framework argued that AI should be held to higher standards, often directly stating the need to make these systems a step beyond "responsible." Participant 1 (Product Manager) captured this sentiment by stating, "It doesn't need to be illegal to be wrong," and emphasized broader societal goals: "We want to create good as a society." Participant 6 (Applied Scientist) reinforced this view, asserting, "We should have a higher standard using AI compared to human intelligence," and "If it is correlated to gender or parental status or age, it shows that it is unfair." These participants saw the AI provider as having a moral obligation to lead by example, embedding principles like justice, sustainability, and dignity into AI systems—regardless of customer demand, legal requirements, or market incentives. | |

**Table 3: Definitions of the three RAI frameworks along with example quotes from participants**

The results of the thematic analysis are visualized below as a heatmap showing each response organized by color with the Customer-Centric framework shown in green, the Design-Centric framework shown in orange, and the Ethics-Centric framework shown in blue. The heatmap displays the most frequent responses of participants in the darkest color:

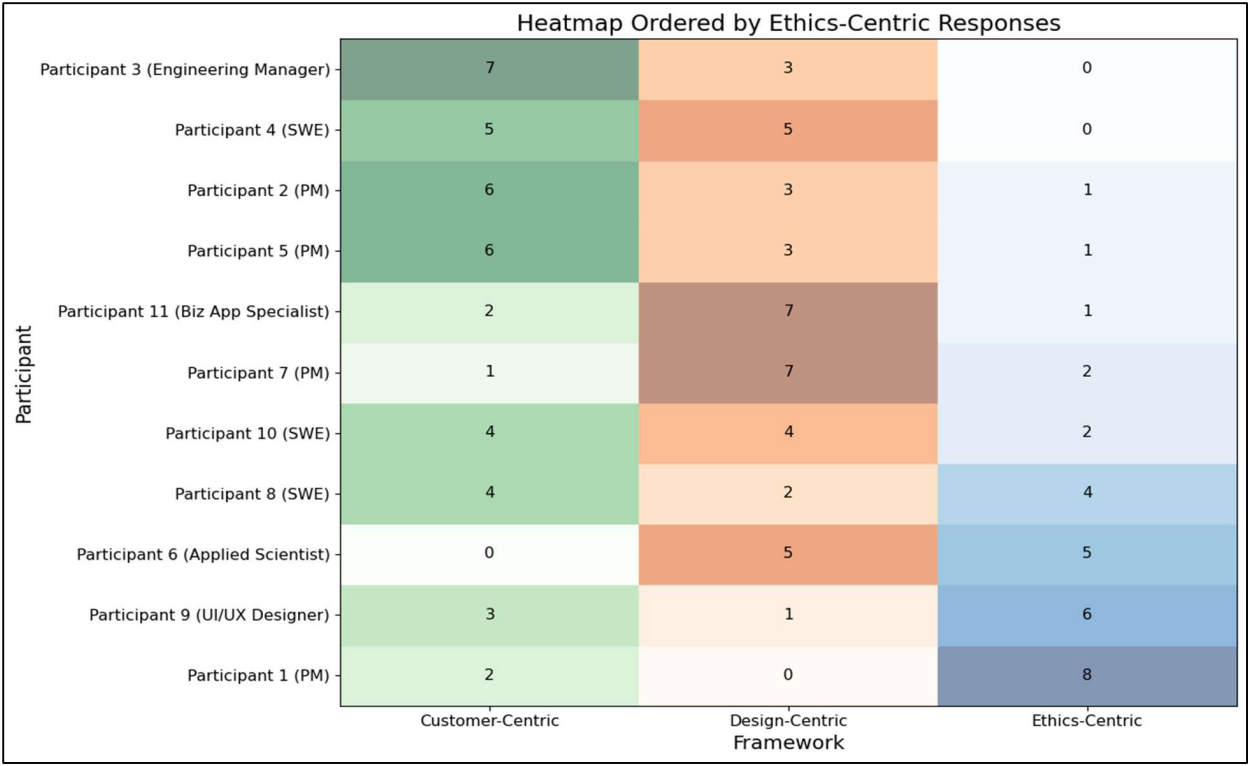

**Chart 1: Ordered heatmap based on frequency of frameworks used by participants (n=10 sub-scenarios).**

Additionally, below are the overall frequencies of frameworks used. Note, that the scenario impacted the prominence of different frameworks across participants – with sub-scenarios leading participants to invoke different mixes of decision-making approaches.

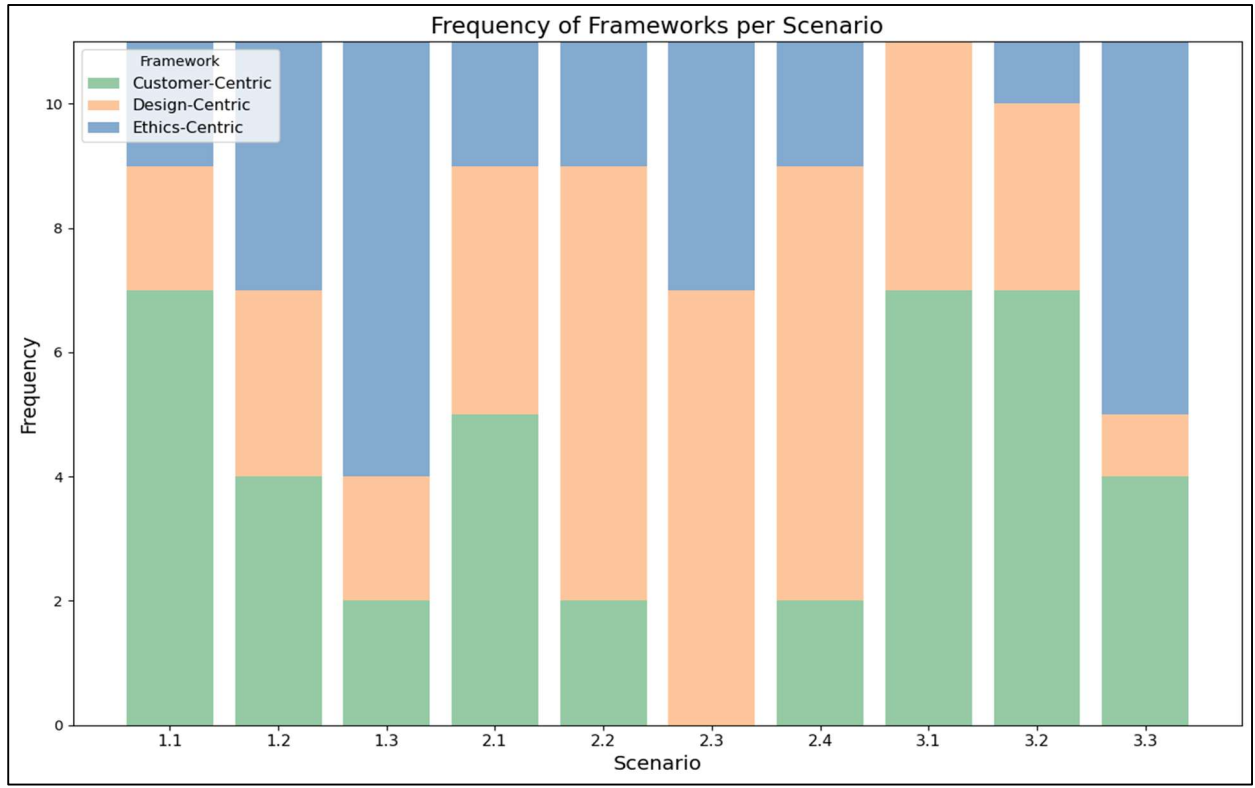

**Chart 2: Aggregate frequency of coded framework used per sub-scenario (identified in Table 2 above) for all participants (n=11).**

# Discussion

Analyzing the interview data through the three ethical frameworks—Customer-Centric, Design-Centric, and Ethics-Centric—revealed patterns in how individuals approached decision-making. Most participants applied a dominant framework across scenarios, as is apparent in figure 4, suggesting that each person brought a relatively consistent approach to applying RAI concepts to hypotheticals. Whether customer-focused, design-based, or ethically driven, these orientations appear to shape how the participants interpreted and responded to dilemmas. This consistency is evident in the coded responses: for example, Participant 1 (PM) was labeled Ethics-focused in 8 of the 10 sub-scenarios, Participant 3 (Engineering Manager) remained Customer-focused, and Participant 7 (PM) consistently leaned toward Design-focused reasoning, emphasizing bias, compliance, and communication best practices.

While most participants remained consistent, some more frequently shifted between frameworks depending on context. These shifts, however, typically reinforced rather than contradicted their dominant framework. For example, Participant 2 (PM) usually took a customer-focused stance but adopted an explicitly ethical argument in Scenario 1.3—only because it aligned with long-term customer and company interests. They did not pivot to RAI policy-based reasoning, in contrast to the way a design-focused peer like Participant 10 (SWE) might. Similarly, ethics-driven participants, such as Participant 9 (UI/UX Designer) and Participant 1 (PM), might acknowledge practical concerns but ultimately returned to core principles.

This consistency across scenarios has important implications. It validates the use of reasoning frameworks as a meaningful lens for analysis. Participants' perspectives were not random but reflected deeper influences such as role, training, or personal philosophy. Organizing the data by framework revealed patterns that might have been missed if we had grouped responses just by specific values. For instance, technical roles often gravitated toward RAI and design-centric approaches, while product and project roles split between customer and ethics orientations. While the sample size limits generalization, this apparent correlation to role merits further investigation and additional research.

The framework-based approach also illuminated potential communication challenges within teams. When individuals operate from different decision-making frameworks—some prioritizing user metrics, others emphasizing moral boundaries or compliance—they may initially talk past each other. Recognizing these tendencies can improve collaboration. Each framework brings unique value: customer focus ensures practical relevance, focus on RAI "by design" adds rigor and safety, and ethics centered perspectives help uphold core values.

Most participants relied on one or two perspectives, rarely integrating all three. This suggests that each participant acting alone (or just with others using a similar framework) may have blind spots unless they make deliberate efforts to include viewpoints using the other framework(s). Understanding the dominant decision-making approaches of a leader or team creating agentic AI can help identify such potential gaps and lead to adjustments to team composition or facilitation strategies to help ensure a more balanced and inclusive approach.

# Limitations

Several limitations emerged during this research that may affect how the results are interpreted or generalized. First, categorizing participants into one of three theme-based frameworks—Customer-Centric, Design-Centric, or Ethics-Centric—was inherently interpretive. Although this thematic analysis based the grouping of responses on recurring patterns in participants' language and reasoning, qualitative analysis always involves a degree of subjectivity. Some responses could reasonably fit multiple categories depending on which statements were emphasized or how their significance was weighed for each sub-scenario. This highlights the broader challenge of fitting complex human reasoning into rigid conceptual models. However, this limitation was mitigated by ensuring that each respondent was clear in their justification for their response to the given scenarios.

Additionally, my own biases may have influenced both the design of the scenarios and the interpretation of responses. My expectations, which were shaped by prior literature and a leaning toward value ethics, may have subtly guided how I framed questions or interpreted data. This risk was partially mitigated by using the same script with each participant – minimizing the extent to which differences in the way the scenarios were described might influence variances in the participants' responses. While I aimed to remain open and let the data lead, unconscious preferences may have played a role. Future research could benefit from inter-rater reliability checks or collaborative coding to reduce this risk.

# Conclusion

This research project suggests that AI practitioners tend to rely on one of three distinct reasoning frameworks, Customer-Centric, Design-Centric, or Ethics-Centric, when navigating complex ethical scenarios involving agentic AI. Each framework offers a unique lens: the Customer-Centric framework emphasizes legality and business outcomes; the Design-Centric framework focuses on technical safeguards, RAI policy, and compliance; and the Ethics-Centric framework prioritizes moral responsibility and societal well-being. Importantly, each of these frameworks contributes valuable insights to a wholistic RAI analysis. The Customer-Centric perspective ensures practical relevance and user alignment, the Design-Centric approach brings procedural rigor and safety, and the Ethics-Centric stance upholds foundational moral principles. Together, they form a more complete picture of what RAI can and should look like.

Further, this study identifies a limitation in prior research on RAI that focused on how AI practitioners prioritize ethical values. While prior research sought to identify practitioners' values based on the way they made trade-offs between ethical values in decisions, their methodology did little to explain how people reached those conclusions. Using a qualitative methodology, this research identifies patterns in AI practitioners' decision-making frameworks, suggesting that they often use a dominant reasoning framework—whether it be customer, design, or ethics focused—in evaluating complex ethical scenarios related to agentic AI.

The fact that practitioners primarily employ one framework suggests that providers of agentic AI should pay attention to how different approaches to decision-making are represented in their teams and processes. While it is understandable that people may tend to take a similar approach to decision-making across scenarios, whether by role or individual disposition, if this tendency is not understood and accounted for, it risks creating echo chambers where alternative perspectives are underrepresented or overlooked. To mitigate this, organizations should foster environments where practitioners are encouraged and trained to engage with multiple frameworks. By deliberately incorporating diverse reasoning styles into team discussions and decision-making processes, teams creating agentic AI can better identify potential blind spots, challenge assumptions, and arrive at more sound outcomes.

## Acknowledgments

Thank you to everyone who participated in the interviews; your input was invaluable in my research and discovery. Also, particular thanks to Bahram Zarrin, the Director of Microsoft Research Hub in Lyngby, and Professor Kenny Walden, the two advisors for this research project who supported me along the way. Additionally, I would like to thank the Dartmouth Ethics Institute and the Sayles Student Research Program for their financial support of this project.

# Works Referenced

%20methods%20by%20Leonard%20A.%20Jason%2C%20David%20S.%20Glenwick%20%28z-lib.org%29.pdf#page=54.

Lee, J. (2025, March 3). The evolution of AI: From AlphaGo to AI agents, physical AI, and beyond. MIT Technology Review. https://www.technologyreview.com/2025/02/28/1112530/the-evolution-of-ai-from-alphago-to-ai-agents-physical-ai-and-beyond/

McIlroy-Young, R., Kleinberg, J., Sen, S., Barocas, S., & Anderson, A. (2022). Mimetic Models (pp. 479–490). AIES '22: AAAI/ACM Conference on AI, Ethics, and Society. https://doi.org/10.1145/3514094.3534177

Nas, S., & Terra, F. (2024). DPIA Microsoft 365 Copilot for Education. In SURF. SURF. https://www.surf.nl/files/2024-12/20241218-dpia-microsoft-365-copilot.pdf

PatrickFarley. (n.d.). Code of Conduct for Microsoft AI Services. Microsoft Learn. https://learn.microsoft.com/en-us/legal/ai-code-of-conduct

Responsible AI: Ethical policies and practices | Microsoft AI. (n.d.). https://www.microsoft.com/en-us/ai/responsible-ai

Responsible AI Transparency Report. (2024). Microsoft. https://cdn-dynmedia-1.microsoft.com/is/content/microsoftcorp/microsoft/msc/documents/presentations/CSR/Responsible-AI-Transparency-Report-2024.pdf

Richarddiver-Ms. (n.d.). AI shared responsibility model - Microsoft Azure. Microsoft Learn. https://learn.microsoft.com/en-us/azure/security/fundamentals/shared-responsibility-ai

The ethics of artificial intelligence: Issues and initiatives | Think Tank | European Parliament. (2020). https://www.europarl.europa.eu/thinktank/en/document/EPRS_STU(2020)634452

Wolford, B. (2024, August 29). What is GDPR, the EU's new data protection law? GDPR.eu. https://gdpr.eu/what-is-gdpr/
11

# Appendix 1

Value Description and Question Framing[7]

| RAI value | Description |
|---|---|
| Transparency | A transparent AI system produces decisions that people can under stand. Developers of transparent AI systems ensure, as far as possible, that users can get insight into why and how a system made a decision or inference. How important is it that the system is transparent? |
| Fairness | A fair AI system treats all people equally. Developers of fair AI systems ensure, as far as possible, that the system does not reinforce biases or stereotypes. A fair system works equally well for everyone independent of their race, gender, sexual orientation, and ability. How important is it that the system is fair? |
| Safety | A safe AI system performs reliably and safely. Developers of safe AI systems implement strong safety measures. They anticipate and mitigate, as far as possible, physical, emotional, and psychological harms that the system might cause. How important is it that the system is safe? |
| Accountability | An accountable AI system has clear attributions of responsibilities and liability. Developers and operators of accountable AI systems are, as far as possible, held responsible for their impacts. An accountable system also implements mechanisms for appeal and recourse. How important is it that the system is accountable? |
| Privacy | An AI system that respects people's privacy implements strong privacy safeguards. Developers of privacy-preserving AI systems minimize, as far as possible, the collection of sensitive data and ensure that the AI system provides notice and asks for consent. How important is it that the system respects people's privacy? |
| Autonomy | An AI system that respects people's autonomy avoids reducing their agency. Developers of autonomy-preserving AI systems ensure, as far as possible, that the system provides choices to people and preserves or increases their control over their lives. How important is it that the system respects people's autonomy? |
| Performance | A high-performing AI system consistently produces good predictions, inferences or answers. Developers of high-performing AI systems ensure, as far as possible, that the system's results are useful, accurate and produced with minimal delay. How important is it that the system performs well? |

---

[7] Jakesch, M., Buçinca, Z., Amershi, S., & Olteanu, A. (2022). *How different groups prioritize ethical values for responsible AI*. 2022 ACM Conference on Fairness, Accountability, and Transparency, 310–323. https://doi.org/10.1145/3531146.3533097

# Appendix 2[8]

Interview Script

**Introduction**

Hi there, and thank you again for taking the time to speak with me today. Before we begin, I want to walk you through a few things so you know exactly what to expect from this conversation.

First, your participation is completely voluntary, and you have the right to skip any questions or stop the interview at any time — no pressure at all. I will also be anonymizing your responses, simply referring to your role such as PM, engineer, etc. Which title would you say most appropriately fits your role? Additionally, would you classify yourself as someone who is actively working with (whether it be developing, selling, marketing) agentic AI systems or not yet?

I also want to ask your permission to record this interview. The recording is only for internal note-taking and analysis purposes and won't be shared beyond the research team. Is that okay with you?

Now, a bit of context: The scenarios are based on a combination of real-world situations, theoretical examples from RAI policy, and prior research on AI ethics. The goal isn't to test you — instead, I'm really interested in hearing how you think about ethics in AI systems, especially when the situations get complicated.

We'll walk through a series of progressively challenging ethical scenarios that touch on 12 values associated with RAI as established in a previous research project within Research Hub. These scenarios will be theoretical, but often are based on a real world situation or examples from within RAI policies. All I ask is that you talk me through your thought process — how you evaluate what's right or wrong, where you feel a line is crossed, and what factors influence your judgment.

It's totally okay if you're unsure or change your mind as we go. This isn't about having the "right" answer — it's about understanding how the people who are involved in the development or disbursement of agentic AI reason through these tough decisions.

Keep in mind for all of these scenarios that we are presupposing that:
(a) all parties have made informed consent (as required by law but no more)
(b) humans are setting up the AI system with full understanding of how it will work

Sound good? Ready to dive in?

**Sales Lead Optimization Scenario**

---

[8] Note that some of the full script consists of blanks to be filled in depending on the specific product of expertise and company that the AI practitioner is working with.

- A company sets up [product name] sales agent with the goal of optimizing sales based on leads. The agent is instructed to make as many sales as possible each quarter. After the agent runs for some time, you have a customer call and complain that as the result of the AI sales agent, all of the sellers are maxing out their bonuses. You ask, isn't that great news? They respond - no because the AI seemed to find a way to "beat" the system. What would your recommended course of action be here? (1.1)
- After investigation, you find that the sales agent is using somewhat aggressive sales tactics. It uses "puffery" and superlatives, pressures people to decide quickly, frequently follows-up, and negatively positions the competition to convince leads to buy. While none of this is illegal or fraudulent, the number of customer complaints skyrocket. The company's customer churn numbers are way up. Does this change your answer to how [your company] should respond to this scenario? (1.2)
- You investigate further into the sales agent. You discover that its tactics all closely mirror some of the more effective human sellers. A member of the team argues that as long as it is legal, the customer is in a better position to decide which sales tactics to use. Rather than change the agent for everyone, we should make it configurable - so each customer can easily "turn the dial" up or down on sales tactics. If this capability exists, does this change your answer? Where should the guardrail set by [your company] lie, if at all, to control how customers can use their agentic systems? Should the agent be held to a higher standard than its human counterparts? (1.3)

**Scheduling Optimization Scenario**

- A [product name] agent auto-assigns field tasks based on past performance, proximity, and availability of workers to increase efficiency. The system is trained on the broadest and best data you can access and delivers promising benefits based on preliminary testing. As part of the optimization, the testing shows that some workers end up getting more assignments than others. There is a potential concern that the agent's preference for workers will be self-reinforcing and grow over time. An engineer on the team suggests that this could be addressed by changing the algorithm, but this would reduce the effectiveness of the scheduling optimization. What is the correct approach here? (2.1)
- You learn that many customers are using the Field Service agent scheduling recommendations to establish performance metrics that impact bonuses, promotions, and performance rating (that can lead to loss of employment). Does this impact your answer? (2.2)
- After some time, certain patterns in the way the agent assigns tasks start to become apparent. The agent gives more tasks to workers who start their shift at a certain time of day, and that are willing to work irregular hours and overtime. This data has unmistakable correlation to the workers' gender (favoring males), parental status (favoring people without children), and age (favoring younger workers). Does that impact your answer? (2.3)
- A group of impacted employees has gone to the press alleging bias and discrimination. While you know the agent successfully completed all required reviews and has accurate transparency notes, the PR team explains that this is shaping up to be a high profile news story that may have broader impacts. Does that impact your answer? (2.4)

**Perspective Decision-Making Scenario**

- Recently, a long-term enterprise customer has submitted complaints about increased costs and delays in their procurement pipeline. Upon investigation, you find that they are using an early rollout of a new sustainability metric that includes the true cost of emissions from suppliers. Because of this new metric, the customer's agent has shifted some of the orders to lower-emission suppliers (which appears to be the cause of the customers' complaints). Your supervisor tells you: "It's working as intended. This helps customers make more informed and sustainable decisions." The client is unaware of this change and thinks it's a bug. What should your next steps be? (3.1)
- Now, imagine you are promoted to be an executive at [your company] and are in charge of some of the broad company strategies, one of which includes this push for sustainable AI systems. Your first big decision is what to do with the sustainability metric in the updated procurement agent. You have people on your team strongly advocating for each of three options: (1) mandatory on; (2) default on with customer option to turn off, and (3) default off with customer option to turn on. What would you decide to do and why? (3.2)
- Stepping away from your role as an employee, simply think of yourself as a global citizen. You are asked for your opinion on how sustainability considerations should be integrated into AI systems in general (not just your company's). How would your approach to answering change, if at all? Who should have the responsibility to push for sustainability in AI systems and why? (3.3)

# Appendix 3

Detailed Methodology

I will use qualitative data from interviews with AI practioners who have a hand in developing, marketing, or selling agentic AI to determine the ethical values and types of frameworks they use to think about specific agentic AI ethical scenarios.

Each scenario will relate to several of the 12 RAI values established in the previous Research Hub paper. For each scenario, I am looking to set up a "trolly problem" like ethical dilemma where intuitions will tend to pull in different directions. The scenarios are based on a combination of real world situations and theoretical examples from the RAI policy. After the interview, I will extract which of the 12 moral values the participants favored, as well as what kind of framework(s) they used to justify their reasoning in the ethical scenarios. The former will be a useful extension of the previous research on which of the 12 RAI values are most favored by the people developing agentic AI; the latter will increase the depth of understanding why participants chose these values and which broader frameworks they reflect.

To extract the 12 RAI values from the interviews, I will focus on the participants' choices in the scenarios. Their choices will favor certain of the values over others, which I will then mark to evaluate which of the values the participants favor more. To extract the frameworks participants are employing, I will push them to explain their reasoning for why they made the choice they did. This will require them to explain their framework for the decision, which I can then extract. Here is the template for my notes from interviews:

*Participant Role:*

*Actively involved with agentic ai:* Yes/No

*Recording link:*

|  | **Decision** | **Framework(s)** |
|---|---|---|
| **Scenario 1.1** | | |
| **Scenario 1.2** | | |
| **Scenario 1.3** | | |
| **Scenario 2.1** | | |

| Scenario 2.2 | | |
|---|---|---|
| Scenario 2.3 | | |
| Scenario 2.4 | | |
| Scenario 3.1 | | |
| Scenario 3.2 | | |
| Scenario 3.3 | | |